\documentclass[12pt]{article}
\usepackage{amssymb,amsfonts}
\usepackage{epsf,epsfig}
\def \nn{{\bf n}}

\begin{document}
\baselineskip 18pt

\title{Free from spurious solutions integral equation for three-magnon bound states in $1D$ $XXZ$ ferromagnet}
\author{P.~N.~Bibikov\\ \it Saint-Petersburg State University,\\
\it 198504 Ulyanovskaya 1, Petrodvorets, Saint-Petersburg, Russia}

\maketitle

\vskip5mm

\begin{abstract}
A new integral equation for three-magnon bound states in XXZ spin chain is suggested. Unlike the one presented by
C. K. Majumdar about 40 years ago this equation does not have spurious solutions. Such an advantage is a result of
decomposition of the wave function in the Bloch basis rather than in the basis of flat waves.
\end{abstract}

\maketitle

\section{Introduction}

An integral equation for three magnon bound states in Heisenberg ferromagnet was first suggested by C. K. Majumdar
\cite{1} and then directly obtained from spin Hamiltonian by several authors \cite{2},\cite{3}.
The main advantage of this approach is its applicability to lattices with dimensions bigger than one.
The main shortcoming is a presence of spurious solutions \cite{3},\cite{4},\cite{5} originated due to slightly
inadequate representation for a three-magnon state
\begin{equation}
|3\rangle=\sum_{\nn_1,\nn_2,\nn_3}u_{\nn_1,\nn_2,\nn_3}{\mathbf S}_{\nn_1}^-{\mathbf S}_{\nn_2}^-
{\mathbf S}_{\nn_3}^-|0\rangle.
\end{equation}
Here $|0\rangle$ is the ferromagnetic ground state (${\mathbf S}_{\nn_j}^+|0\rangle=0$ and the indices
$\nn_j$ numerate a $D$-dimensional lattice. Amplitudes $u_{\nn_1,\nn_2,\nn_3}$ with $\nn_j=\nn_k$ for $j\neq k$ are
unphysical and are separated in the ${\rm Shr\ddot odinger}$ equation from the physical ones \cite{2},\cite{3}.
It is unphysical amplitudes that spurious solutions originate from. They have already been classified \cite{6}.
In the present paper for the case $D=1$ we represent the three-magnon bound state in the form which contains only 
physical amplitudes and then obtain the corresponding integral equation.

\section{Hamiltonian and ${\rm\bf Shr\ddot odinger}$ equation}

Hamiltonian of the $S=1/2$ $XXZ$  ferromagnetic chain
\begin{equation}
\hat H=-\sum_n\Big[\frac{1}{2}\Big({\mathbf S}_n^+{\mathbf S}_{n+1}^-+{\mathbf S}_n^-{\mathbf S}_{n+1}^+\Big)
+\Delta\Big({\mathbf S}_n^z{\mathbf S}_{n+1}^z-\frac{1}{4}\Big)\Big].
\end{equation}
acts in an infinite tensor product of ${\mathbb C}^2$ subspaces associated with each site of the chain. Here
${\bf S}_n^{j}$ are the $S=1/2$ spin operators corresponding to $n$-th site.
For a three-magnon state we suggest the following representation
\begin{equation}
|3\rangle=\sum_{m<n<p}{\rm e}^{i(m+n+p)k/3}b_{n-m,p-n}(k){\mathbf S}_m^-{\mathbf S}_n^-{\mathbf S}_p^-|0\rangle.
\end{equation}
where
\begin{equation}
|0\rangle=\prod_n\otimes|\uparrow\rangle_n,
\end{equation}
is the ferromagnetic ground state.

It is convenient to represent the corresponding ${\rm Shr\ddot odinger}$ equation in the form
\begin{equation}
\Big([H_0(k)+\Delta(3+V)]b(k)\Big)_{m,n}=Eb_{m,n}(k),
\end{equation}
where
\begin{eqnarray}
\Big(H_0(k)b(k)\Big)_{m,n}&=&\frac{{\rm e}^{ik/3}}{2}\Big((\delta_{m,1}-1)b_{m-1,n}(k)+
(\delta_{n,1}-1)b_{m+1,n-1}(k)-b_{m,n+1}(k)\Big)\nonumber\\
&+&\frac{{\rm e}^{-ik/3}}{2}\Big((\delta_{m,1}-1)b_{m-1,n+1}(k)+(\delta_{n,1}-1)b_{m,n-1}(k)
-b_{m+1,n}(k)\Big),\nonumber\\
\Big(Vb(k)\Big)_{m,n}&=&-\delta_{m,1}b_{1,n}(k)-\delta_{n,1}b_{m,1}(k).
\end{eqnarray}

\section{Three-magnon scattering states at ${\bf\Delta=0}$}

In the free ($\Delta=0$) case Eq. (5) has the following system of $k$-independent Bloch solutions
\begin{eqnarray}
b^{(0)}_{m,n}({\bf q})=\frac{1}{\sqrt{6}}\Big({\rm e}^{i[q_1m+q_2n]}-{\rm
e}^{i[q_1m+(q_1-q_2)n]}
+{\rm e}^{i[(q_2-q_1)m-q_1n]}\nonumber\\
-{\rm
e}^{i[(q_2-q_1)m+q_2n]}
+{\rm e}^{-i[q_2m+(q_2-q_1)n]}-{\rm e}^{-i[q_2m+q_1n]}\Big),\quad {\bf q}\equiv(q_1,q_2),
\end{eqnarray}
with dispersion
\begin{equation}
E_0(k,{\bf q})=-\cos{\left(\frac{k}{3}-q_1\right)}-\cos{\left(\frac{k}{3}+q_1-q_2\right)}-
\cos{\left(\frac{k}{3}+q_2\right)}.
\end{equation}

It is convenient to represent the wave function (7) in a compact form
\begin{equation}
b^{(0)}_{m,n}({\bf q})=\frac{1}{\sqrt{6}}\sum_{\omega\in{\cal S}_3}(-1)^{\deg{\omega}}{\rm e}^{i[\omega^{(1)}(q_1)m+
\omega^{(2)}(q_2)n]},\quad (-1)^{\deg{\omega_{jkl}}}=\varepsilon_{jkl}.
\end{equation}
where $\omega$ give the following representations of ${\cal S}_3$ (the permutation group of three elements)
\begin{eqnarray}
\omega_{123}(q_1,q_2)=(q_1,q_2),\quad
\omega_{132}(q_1,q_2)=(q_1,q_1-q_2),\quad
\omega_{213}(q_1,q_2)=(q_2-q_1,q_2),\quad\nonumber\\
\omega_{231}(q_1,q_2)=(q_2-q_1,-q_1),\quad
\omega_{312}(q_1,q_2)=(-q_2,q_1-q_2),\quad
\omega_{321}(q_1,q_2)=(-q_2,-q_1).\quad
\end{eqnarray}
From (9) follows the symmetry property
\begin{equation}
b^{(0)}_{m,n}(\omega({\bf q}))=(-1)^{\deg{\omega}}b^{(0)}_{m,n}({\bf q}),
\end{equation}
which reduces $\Omega$, the fundamental region for the parameters $q_1$ and $q_2$. If
\begin{equation}
\Omega_0:\quad0\leq q_1<2\pi,\quad0\leq q_2<2\pi,
\end{equation}
then
\begin{equation}
\Omega_0=\cup_{\omega\in{\cal S}_3}\omega(\Omega).
\end{equation}

The system of functions (9) is normalized and complete
\begin{eqnarray}
\sum_{m,n=1}^{\infty}\bar b^{(0)}_{m,n}({\bf q})b^{(0)}_{m,n}({\bf\tilde q})&=&
\delta(q_1-\tilde q_1)\delta(q_2-\tilde q_2),\\
\frac{3}{2\pi^2}\int_{\Omega}\bar b^{(0)}_{\tilde m,\tilde n}({\bf q})b^{(0)}_{m,n}({\bf q})dq_1dq_2&=&
\frac{1}{(2\pi)^2}\int_0^{2\pi}dq_1\int_0^{2\pi}dq_2\bar b^{(0)}_{\tilde m,\tilde n}({\bf q})b^{(0)}_{m,n}({\bf q})=
\delta_{m\tilde m}\delta_{n\tilde n}.
\end{eqnarray}

In order to prove the normalization condition
one should represent the product of wave functions in the sum (14) as follows
\begin{eqnarray}
\bar b^{(0)}_{m,n}({\bf q})b^{(0)}_{m,n}({\bf\tilde q})&=&
\frac{1}{6}\sum_{\omega,\tilde\omega\in{\cal S}_3}(-1)^{\deg{\omega}+\deg{\tilde\omega}}
{\rm e}^{-i[(\omega^{(1)}(q_1)-\tilde\omega^{(1)}(\tilde q_1))m+(\omega^{(2)}(q_2)-
\tilde\omega^{(2)}(\tilde q_2))n]}\nonumber\\
&=&\sum_{\omega\in{\cal S}_3}(-1)^{\deg{\omega}}g_{m,n}({\bf q}-\omega({\bf\tilde q})),
\end{eqnarray}
where
\begin{eqnarray}
g_{m,n}({\bf q})&=&\frac{1}{6}\sum_{\omega\in{\cal S}_3}{\rm e}^{-i[\omega^{(1)}(q_1)m+
\omega^{(2)}(q_2)n]}=\frac{1}{6}\Big({\rm e}^{-i(q_1m+q_2n)}
+{\rm e}^{-i[q_1(m+n)-q_2n]}\nonumber\\
&+&{\rm e}^{-i(q_2m-q_1(m+n)]}
+{\rm e}^{i[q_1m-q_2(m+n)]}
+{\rm e}^{i[q_2(m+n)-q_1n]}+{\rm e}^{i[q_2m+q_1n]}\Big).
\end{eqnarray}
Since for each $d_{m,n}$
\begin{equation}
\sum_{m,n>0}(d_{m,m+n}+d_{m+n,n})=\sum_{m,n>0}d_{m,n}-\sum_{n>0}d_{n,n},
\end{equation}
one readily obtains from (17)
\begin{eqnarray}
\sum_{m,n>0}g_{m,n}({\bf q})&=&\frac{1}{6}\Big[\sum_{m,n=-\infty}^{\infty}{\rm e}^{i(q_1m+q_2n)}-
\sum_{n=-\infty}^{\infty}\Big({\rm e}^{i(q_1+q_2)n}+{\rm e}^{i(q_1-q_2)n}\Big)+1\Big]\nonumber\\
&=&\frac{1}{6}\Big[4\pi^2\delta(q_1)\delta(q_2)-2\pi\Big(\delta(q_1+q_2)+\delta(q_1-q_2)\Big)+1\Big].
\end{eqnarray}
Hence according to (16) and (19)
\begin{eqnarray}
\sum_{m,n>0}\bar b^{(0)}_{m,n}({\bf q})b^{(0)}_{m,n}({\bf\tilde q})&=&\frac{2\pi^2}{3}
\sum_{\omega\in{\cal S}_3}(-1)^{\deg{(\omega)}}\delta(q_1-\omega^{(1)}(\tilde q_1))
\delta(q_2-\omega^{(2)}(\tilde q_2)),
\end{eqnarray}
(other terms cancelled after averaging over ${\cal S}_3$).
But $\omega({\bf\tilde q})\in\Omega$ only for $\omega=\omega_{123}$. So all terms in (20) with $\omega\neq\omega_{123}$
should be omitted and we get Eq. (14).

In order to prove completeness let us first define an action of ${\cal S}_3$ on the configuration space by
the following formula
\begin{equation}
\omega^{(1)}(q_1)m+\omega^{(2)}(q_2)n=q_1\omega^{(1)}(m)+q_2\omega^{(2)}(n).
\end{equation}
According to (10) and (21)
\begin{eqnarray}
\omega_{123}(m,n)=(m,n),\quad\omega_{132}(m,n)=(m-n,-n),\quad\omega_{213}(m,n)=(-m,m+n),\nonumber\\
\omega_{231}(m,n)=(-m-n,m),\quad\omega_{312}(m,n)=(n,-m-n),\quad\omega_{321}(m,n)=(-n,-m).
\end{eqnarray}

In this notations
\begin{equation}
\bar b^{(0)}_{\tilde m,\tilde n}({\bf q})b^{(0)}_{m,n}({\bf q})=
\frac{1}{6}\sum_{\omega,\tilde\omega\in S_3}(-1)^{\deg\omega+\deg\tilde\omega}
{\rm e}^{i[q_1(\omega^{(1)}(m)-\tilde\omega^{(1)}(\tilde m))+q_2(\omega^{(2)}(n)-\tilde\omega^{(2)}(\tilde n))]}
\end{equation}
So
\begin{equation}
\frac{1}{(2\pi)^2}\int_0^{2\pi}dq_1\int_0^{2\pi}dq_2\bar b^{(0)}_{\tilde m,\tilde n}({\bf q})b^{(0)}_{m,n}({\bf q})=
\sum_{\omega\in S_3}(-1)^{\deg\omega}\delta_{m,\omega^{(1)}(\tilde m)}\delta_{n,\omega^{(2)}(\tilde n)}.
\end{equation}
Since $m,n,\tilde m,\tilde n>0$ Eq. (15) follows from (22) and (24).

\section{${\rm\bf Shr\ddot odinger}$ equation in the $\bf q$-space}

Let
\begin{equation}
b_{m,n}(k)=\frac{3}{2\pi^2}\int_{\Omega}b^{(0)}_{m,n}({\bf q})b(k,{\bf q})d{\bf q},
\end{equation}
or equivalently
\begin{equation}
b(k,{\bf q})=\sum_{m,n>0}\bar b^{(0)}_{m,n}({\bf q})b_{m,n}(k),
\end{equation}
be a decomposition of the wave function in the basis of Bloch functions. In this framework Eq. (5) takes the form
\begin{equation}
(E_0(k,{\bf q})+3\Delta-E)b(k,{\bf q})=-6\Delta\int_{\Omega}V({\bf q},{\bf\tilde q})b(k,{\bf\tilde q})d{\bf\tilde q},
\end{equation}
where
\begin{eqnarray}
V({\bf q},{\bf\tilde q})&=&-\frac{1}{4\pi^2}\sum_{n=1}^{\infty}\Big(\bar b^{(0)}_{1,n}({\bf q})b_{1,n}({\bf\tilde q})
+\bar b^{(0)}_{n,1}({\bf q})b^{(0)}_{n,1}({\bf\tilde q})\Big).
\end{eqnarray}
As it follows from (11), (25) and (28)
\begin{equation}
b(k,\omega({\bf q}))=(-1)^{\deg\omega}b(k,{\bf q}),\quad V(\omega({\bf q}),{\bf\tilde q})=
V({\bf q},\omega({\bf\tilde q}))=(-1)^{\deg\omega}V({\bf q},{\bf\tilde q}).
\end{equation}
Hence the integration over $\Omega$ in Eq. (27) may be extended to integration over $\Omega_0$
\begin{equation}
(E_0(k,{\bf q})+3\Delta-E(k))b(k,{\bf q})=-\Delta\int_0^{2\pi}d\tilde q_1
\int_0^{2\pi}d\tilde q_2V({\bf q},{\bf\tilde q})
b(k,{\bf\tilde q}),\quad{\bf q}\in\Omega_0.
\end{equation}

For scattering states
\begin{equation}
E_{scatt}(k,{\bf q})=E_0(k,{\bf q})+3\Delta,
\end{equation}
and hence the right side of (30) should be zero. For bound states we assume
\begin{equation}
E_{bound}(k)<{\rm min}_{\bf q}(E_{scatt}(k,{\bf q}))=3\Delta-3\cos{\frac{k}{3}}.
\end{equation}

Let us now calculate $V({\bf q},{\bf\tilde q})$. According to (16) and (28)
\begin{equation}
V({\bf q},{\bf\tilde q})=-\frac{1}{4\pi^2}\sum_{\omega\in{\cal S}_3}(-1)^{\deg\omega}\sum_{n=1}^{\infty}
g_{1,n}(\omega({\bf q})-{\bf\tilde q})+g_{n,1}(\omega({\bf q})-{\bf\tilde q}).
\end{equation}

As it follows from (17)
\begin{eqnarray}
\sum_{n=1}^{\infty}g_{1,n}({\bf q})+g_{n,1}({\bf q})=\frac{1}{6}\sum_{n=1}^{\infty}\Big[{\rm e}^{-iq_1}\Big(
{\rm e}^{-iq_2n}+{\rm e}^{iq_2(n+1)}+{\rm e}^{i(q_2-q_1)n}+{\rm e}^{i(q_1-q_2)(n+1)}\Big)\nonumber\\
+{\rm e}^{i(q_1-q_2)}\Big({\rm e}^{iq_1n}+{\rm e}^{-iq_1(n+1)}+{\rm e}^{-iq_2n}+{\rm e}^{q_2(n+1)}\Big)
+{\rm e}^{iq_2}\Big({\rm e}^{i(q_2-q_1)n}+{\rm e}^{i(q_1-q_2)(n+1)}\nonumber\\
+{\rm e}^{iq_1n}+{\rm e}^{-iq_1(n+1)}\Big)\Big]
=\frac{\pi}{3}\sum_{\omega\in{\cal S}_3}{\rm e}^{i\omega^{(1)}(q_1)}\delta(\omega^{(2)}(q_2))
-\frac{2}{3}\Big(\cos{q_1}+\cos{q_2}+\cos{(q_1-q_2)}\Big).
\end{eqnarray}
Eqs. (33) and (34) result in
\begin{equation}
V({\bf q},{\bf\tilde q})
=-\frac{1}{12\pi}\sum_{\omega,\tilde\omega\in{\cal S}_3}(-1)^{\deg{\omega}-\deg{\tilde\omega}}
{\rm e}^{i[\omega^{(1)}(q_1)-\tilde\omega^{(1)}(\tilde q_1)]}\delta[\omega^{(2)}(q_2)-\tilde\omega^{(2)}(\tilde q_2)].
\end{equation}

A substitution of (35) into (30) and evaluation of the sum over $\tilde\omega$ using appropriate changes of
${\bf\tilde q}$ gives for a bound state
\begin{eqnarray}
(E_0(k,{\bf q})+3\Delta-E_{bound}(k))b(k,{\bf q})&=&\frac{\Delta}{2\pi}\int_0^{2\pi}d\tilde q_1
\int_0^{2\pi}d\tilde q_2\sum_{\omega\in{\cal S}_3}(-1)^{\deg{\omega}}\nonumber\\
&&\cdot{\rm e}^{i[\omega^{(1)}(q_1)-\tilde q_1]}\delta[\omega^{(2)}(q_2)-\tilde q_2]b(k,{\bf\tilde q}),
\end{eqnarray}
or equivalently
\begin{equation}
(E_0(k,{\bf q})+3\Delta-E_{bound}(k))b(k,{\bf q})=\Delta\sum_{\omega\in{\cal S}_3}(-1)^{\deg{\omega}}
{\rm e}^{i\omega^{(1)}(q_1)}\psi(k,\omega^{(2)}(q_2)),
\end{equation}
where
\begin{equation}
\psi(k,q)=\frac{1}{2\pi}\int_0^{2\pi}d\tilde q{\rm e}^{-i\tilde q}b(k,\tilde q,q).
\end{equation}
Eqs. (37) and (38) result in an integral equation
\begin{equation}
\psi(k,q)=\frac{\Delta}{2\pi}\sum_{\omega\in{\cal S}_3}(-1)^{\deg{\omega}}
\int_0^{2\pi}\frac{{\rm e}^{i(\omega^{(1)}(\tilde q)-\tilde q)}}{E_0(k,\tilde q,q)+3\Delta-E_{bound}(k)}
\psi(k,\omega^{(2)}(q))d\tilde q,
\end{equation}
or in an extended form
\begin{eqnarray}
\psi(k,q)&=&\frac{\Delta}{2\pi}\int_0^{2\pi}\frac{d\tilde q}{E_0(k,\tilde q,q)+3\Delta-E_{bound}(k)}
\Big[\Big(1-{\rm e}^{i(q-2\tilde q)}\Big)\psi(k,q)\nonumber\\
&+&\Big({\rm e}^{i(q-2\tilde q)}-{\rm e}^{-i(q+\tilde q)}\Big)\psi(k,-\tilde q)
+\Big({\rm e}^{-i(q+\tilde q)}-1\Big)\psi(k,\tilde q-q)\Big].
\end{eqnarray}
For future simplication we notice that
\begin{equation}
E_0(k,\tilde q,q)+3\Delta-E_{bound}(k)=
\frac{\gamma(k,q)(\tilde z-z_+)(\tilde z-z_-)}{\tilde z(z_--z_+)},
\end{equation}
where
\begin{equation}
\tilde z={\rm e}^{i\tilde q},\quad z_{\pm}=\frac{
3\Delta-E_{bound}(k)-\cos{(k/3+q)}\pm\gamma(k,q)}
{\left({\rm e}^{i(k/3-q)}+{\rm e}^{-ik/3}\right)}.
\end{equation}
and
\begin{equation}
\gamma(k,q)=\sqrt{\Big(\cos{\Big(\frac{k}{3}+q\Big)}+E_{bound}(k)-3\Delta\Big)^2
-\Big|{\rm e}^{i(k/3-q)}+{\rm e}^{-ik/3}\Big|^2}.
\end{equation}
From (42), (43) and (32) follows that 
\begin{equation}
z_+\bar z_-=\bar z_+z_-=1,\qquad |z_+|>1,\quad |z_-|<1.
\end{equation}
Hence from (41) and (44) one readily gets
\begin{eqnarray}
\frac{1}{2\pi}\int_0^{2\pi}\frac{(1-{\rm e}^{i(q-2\tilde q)})d\tilde q}{E_0(k,\tilde q,q)+3\Delta-E_{bound}(k)}=
\frac{1-{\rm e}^{iq}\bar z_-^2}{\gamma(k,q)}.
\end{eqnarray}

Now using (45) and the following formulas
\begin{equation}
E_0(k,-\tilde q,q)=E_0(k,q+\tilde q,q)=E_0(k,-q,\tilde q),
\end{equation}
one may reduce (40) to the form
\begin{equation}
\left(1-\frac{\Delta(1-{\rm e}^{iq}\bar z_-^2)}{\gamma(k,q)}\right)\psi(k,q)=
\frac{\Delta}{2\pi}\int_0^{2\pi}\frac{[{\rm e}^{i(q+2\tilde q)}-{\rm e}^{i(\tilde q-q)}+{\rm e}^{-i(\tilde q+2q)}-1]
\psi(k,\tilde q)d\tilde q}{E_0(k,-q,\tilde q)+3\Delta-E_{bound}(k)},
\end{equation}
or
\begin{equation}
\left(1-\frac{\Delta(1-{\rm e}^{iq}\bar z_-^2)}{\gamma(k,q)}\right)\varphi(k,q)=\frac{\Delta}{\pi}\int_0^{2\pi}
\frac{[\cos{3(q+\tilde q)/2}-\cos{(q-\tilde q)/2}]\varphi(k,\tilde q)d\tilde q}
{E_0(k,-q,\tilde q)+3\Delta-E_{bound}(k)},
\end{equation}
where
\begin{equation}
\varphi(k,q)={\rm e}^{iq/2}\psi(k,q).
\end{equation}

Eq. (48) is the main result of the paper. Although it is analogous to Eq. (91) of Ref. 1, its derivation
does not need introduction of any additional constructions (as it has been for the Majumdar equation \cite{2},\cite{3}).
In the Bloch basis Eq. (48) directly follows from representation (3).
According to (26), (38) and (49) up to a normalization constant
\begin{equation}
\varphi(k,q)=\sum_{n=1}^{\infty}b_{n,1}(k){\rm e}^{iq(n+1/2)}+b_{1,n}(k){\rm e}^{-iq(n+1/2)}.
\end{equation}

\section{Checking on the Bethe Ansatz result}

An exact form of the three-magnon wave function is well known \cite{7}. Namely
\begin{equation}
b_{m,n}(k)=z_1^{m-1}(k)z_2^{n-1}(k),
\end{equation}
where
\begin{equation}
z_1(k)=\frac{1}{4\Delta^2-1}\Big(2\Delta{\rm e}^{ik/3}+{\rm e}^{-2ik/3}\Big),\quad
z_2(k)=\bar z_1(k)=\frac{1}{4\Delta^2-1}\Big(2\Delta{\rm e}^{-ik/3}+{\rm e}^{2ik/3}\Big).
\end{equation}
In fact it may be readily proved that (51) satisfies the system (6) with
\begin{eqnarray}
E_{bound}(k)&=&3\Delta-\frac{{\rm e}^{-ik/3}}{2}\left(z_1(k)+\frac{1}{z_2(k)}+\frac{z_2(k)}{z_1(k)}\right)
-\frac{{\rm e}^{ik/3}}{2}\left(z_2(k)+\frac{1}{z_1(k)}+\frac{z_1(k)}{z_2(k)}\right)\nonumber\\
&=&3\Delta-\frac{8\Delta^3+\cos{k}}{4\Delta^2-1}.
\end{eqnarray}

According to (50) and (51)
\begin{equation}
\varphi(k,q)=\frac{{\rm e}^{iq/2}}{{\rm e}^{-iq}-z_1(k)}+\frac{{\rm e}^{-iq/2}}{{\rm e}^{iq}-z_2(k)}.
\end{equation}
Since
\begin{equation}
E_0(k,-q,\tilde q)+3\Delta-E_{bound}(k)=
\frac{\gamma(k,q)(\tilde z-\bar z_+)(\tilde z-\bar z_-)}{\tilde z(\bar z_--\bar z_+)},
\end{equation}
Eq. (48) takes the form
\begin{eqnarray}
\Big(\gamma(k,q)+\Delta({\rm e}^{iq}\bar z_-^2-1)\Big)\left(\frac{{\rm e}^{iq/2}}{{\rm e}^{-iq}-z_1(k)}+
\frac{{\rm e}^{-iq/2}}{{\rm e}^{iq}-z_2(k)}\right)=\frac{\Delta(\bar z_--\bar z_+)}{2\pi i}
\oint\Big({\rm e}^{3iq/2}\tilde z^2\nonumber\\
-{\rm e}^{-iq/2}{\tilde z}-{\rm e}^{iq/2}+{\rm e}^{-3iq/2}\frac{1}{\tilde z}\Big)
\left(\frac{\tilde z}{1-\tilde zz_1(k)}+\frac{1}{\tilde z(\tilde z-z_2(k))}\right)
\frac{d\tilde z}{(\tilde z-\bar z_+)(\tilde z-\bar z_-)}.
\end{eqnarray}

Integral in the right side of (56) may be decomposed and readily calculated
\begin{eqnarray}
\frac{\bar z_--\bar z_+}{2\pi i}\oint\Big({\rm e}^{3iq/2}\tilde z^3-{\rm e}^{-iq/2}{\tilde z}^2-{\rm e}^{iq/2}\tilde z+
{\rm e}^{-3iq/2}\Big)\frac{d\tilde z}{(1-\tilde zz_1(k))(\tilde z-\bar z_+)(\tilde z-\bar z_-)}\nonumber\\
=\left({\rm e}^{3iq/2}\bar z_-^3-{\rm e}^{-iq/2}\bar z_-^2-{\rm e}^{iq/2}\bar z_-+{\rm e}^{-3iq/2}\right)
\frac{1}{1-\bar z_-z_1(k)},
\end{eqnarray}
and ($w=1/\tilde z$)
\begin{eqnarray}
\frac{\bar z_--\bar z_+}{2\pi i}\oint\left({\rm e}^{3iq/2}\tilde z-{\rm e}^{-iq/2}-{\rm e}^{iq/2}\frac{1}{\tilde z}+
{\rm e}^{-3iq/2}\frac{1}{\tilde z^2}\right)
\frac{d\tilde z}{(\tilde z-z_2(k))(\tilde z-\bar z_+)(\tilde z-\bar z_-)}\nonumber\\
=\frac{\bar z_--\bar z_+}{2\pi i}\oint\left({\rm e}^{3iq/2}-{\rm e}^{-iq/2}w-{\rm e}^{iq/2}w^2+{\rm e}^{-3iq/2}w^3\right)
\frac{dw}{(1-wz_2(k))(1-\bar z_+w)(1-\bar z_-w)}\nonumber\\
=\left({\rm e}^{3iq/2}-{\rm e}^{-iq/2}\frac{1}{\bar z_+}-{\rm e}^{iq/2}\frac{1}{\bar z_+^2}+{\rm e}^{-3iq/2}
\frac{1}{\bar z_+^3}\right)\frac{\bar z_+}{\bar z_+-z_2(k)},
\end{eqnarray}
Using (57), (58) and (44) one may reduce Eq. (56) to the form
\begin{eqnarray}
\left(\frac{\gamma(k,q)}{\Delta}+{\rm e}^{iq}\bar z_-^2-1\right)\left(\frac{{\rm e}^{iq/2}}{{\rm e}^{-iq}-z_1(k)}+
\frac{{\rm e}^{-iq/2}}{{\rm e}^{iq}-z_2(k)}\right)=
\frac{{\rm e}^{-3iq/2}({\rm e}^{iq}\bar z_-^2-1)({\rm e}^{2iq}\bar z_--1)}{1-\bar z_-z_1(k)}\nonumber\\
+\frac{{\rm e}^{3iq/2}({\rm e}^{-iq}z_-^2-1)({\rm e}^{-2iq}z_--1)}{1-z_-z_2(k)},
\end{eqnarray}
which may be checked directly by rather cumbersome calculation.

\section{Conclusions}

In the present paper we obtained an integral equation (48) for three-magnon bound states in 1D Heisenberg ferromagnet.
The suggested equation is based on the representation (3) which does not contain unphysical amplitudes related to
spurious solutions. The derivation is based on the decomposition (25) of the wave function in the basis of Bloch wave 
functions. The obtained equation was directly tested on the Bethe Ansatz solution (51). Basing on this result we suggest 
that for a study of bound states in a complex model it is better to decompose a wave function not in the flat waves basis
but in a basis of states related to an appropriate solvable model.

The author is very grateful to P. P. Kulish for a formulation of the problem and to M. I. Vyazovsky for
an interest to the work.

\end{document}